\documentstyle[12pt]{article}
\begin{document}
\begin{center}

{\LARGE \bf Relativistic two-body system }\\
{\LARGE \bf in (1+1)-dimensions}\\

\vspace{2 cm}

{\bf Norman Dombey}\\

\vspace{1 cm}

{\it Center for Theoretical Physics, University of Sussex,\\
Falmer, Brighton BN1 9QH, UK}\\

\vspace{1 cm}

{\bf and}\\

\vspace{1 cm}

{\bf Fuad M. Saradzhev}\\

\vspace{1 cm}

{\it Institute of Physics, Academy of Sciences of Azerbaijan,\\
H.Javid pr. 33, 370143 Baku, Azerbaijan}\\

\vspace{2 cm}

{\bf Abstract}

\end{center}

\rm

The relativistic two-body system in $(1+1)$-dimensional 
quantum electrodynamics is studied . It is proved that the
eigenvalue problem for the two-body Hamiltonian 
without the self-interaction terms reduces to the
problem of solving an one-dimensional stationary Schr\"odinger
type equation with an energy-dependent effective potential which
includes the $\delta$-functional and inverted oscillator parts.
The conditions determining the metastable energy spectrum
are derived, and the energies and widths of the metastable levels
are estimated in the limit of large particle masses. The effects
of the self-interaction are discussed.

\newpage

\begin{center}

\section{INTRODUCTION}
\label{sec: intro}

\end{center}

In the study of two-body systems in quantum theory we often use 
single-particle equations. For hydrogenlike systems, for instance,
we assume that one of the particles is much heavier (proton) and
then reduce the two-body problem to the problem of motion of the
lighter particle (electron) in an external field of the heavier one.
To get more exact solution we need to take into account
two-body effects and start with a two-body equation.

As shown in \cite{gal88}, in $(1+1)$-dimensions the single-particle
Dirac equation allows no hydrogen. The equation has solutions for
a continuous set of energies, and the probability of finding the
electron infinitely separated from the proton remains finite at
all times. Despite the attractive force, the electron and proton
are not confined in a hydrogenlike system with discrete energy
levels. That happens not only for hydrogen atoms with an infinitely
heavy source of potential, but also for positroniumlike systems.

In the present paper, we want to clarify to what extent the
two-body effects influence the result of \cite{gal88}. We aim
to study a relativistic two-body system in $(1+1)$-dimensional
quantum electrodynamics (QED) by making use of a two-body
Dirac equation.  Models in $(1+1)$-dimensions are known to be
useful as simpler models for discussion of many-body aspects
of particle physics, in particular, spontaneous positron
production by supercritical potentials \cite{cal96,dom99}.

To describe two-body systems we usually introduce a composite
field, and there are two ways of deriving equations on this field.
If we rewrite the action of the two-body system entirely in terms
of the composite field, then we can require the action to be
stationary with respect to the variations of this field only.
This way leads to a single two-body equation \cite{bar90} -
\cite{unal86}. However, if we first vary the action with
respect to the individual fields, then we come to a pair of
coupled equations on the composite field. The pair of Dirac
equations formulation of the two-body problem was given in
\cite{crat84,saz86} in the framework of the constrait approach.
The main difference between the two ways is in the role of 
the relative energy ( or its conjugate variable, the relative time).
While in the first way the relative energy drops out of the
two-body equation automatically, in the pair of Dirac equations
formulation it is eliminated by using the compatibility condition
of the two equations and a special choice of the interaction
potential. 

In our paper we follow the single two-body equation formulation
and work in the first-quantized version of QED when both matter
and electromagnetic fields are not quantized. We consider a
system of two massive Dirac fields minimally coupled to a $U(1)$
or electromagnetic field. The electromagnetic field has no
separate local degrees of freedom and can be eliminated between
the coupled Maxwell-Dirac equations, but then nonlinear self-field
terms must be included. In Sect. 2, we derive a relativictic
two-body equation in the self-field ${\rm QED}_{1+1}$ defined
on the line and give the Hamiltonian form of this equation.
In Sect. 3, we find the eigenfunctions and the spectrum of the
two-body Hamiltonian. We study in detail two cases: i) free
motion; ii) the Coulomb interaction, and discuss the effects
of the self-interaction. Sect. 4 contains our conclusions.

\newcommand{\ren}{\renewcommand{\theequation}{2.\arabic{equation}}}
\newcommand{\new}[1]{\renewcommand{\theequation}{2.\arabic{equation}#1}}
\newcommand{\add}{\addtocounter{equation}{-1}}

\begin{center}

\section{TWO-BODY EQUATION}

\end{center}

For our system, the action is
\ren
\begin{equation}
{\rm W}=\int_{- \infty}^{\infty} dt \int_{- \infty}^{\infty} dx
{\cal L}(x,t),
\label{eq: dvaodin}
\end{equation}
\[
{\cal L} = \sum_{k=1}^{2} [ \overline{\psi}_k {\gamma}^{\mu}
(i{\hbar}c {\partial}_{\mu} -e_k A_{\mu}) {\psi}_k  
- m_k c^2 \overline{\psi}_k
{\psi}_k] - \frac{1}{4} F_{\mu \nu} F^{\mu \nu},
\]
where $(\mu,\nu=\overline{0,1})$, ${\gamma}^0=-i{\sigma}_2$,
${\gamma}^0 {\gamma}^1 ={\gamma}^5 ={\sigma}_3$, ${\sigma}_i$
$(i=\overline{1,3})$ are Pauli matrices. The fields ${\psi}_k$ are
two-component Dirac spinors, and $\bar{\psi}_k= {\psi}_k^{\star}
{\gamma}^0$. The partial derivatives are defined as
${\partial}_0 = {\partial}/c{\partial t}$,
${\partial}_1 = {\partial}/{\partial x}$.

The electromagnetic field equations deduced from the action 
(\ref{eq: dvaodin}) are
\begin{equation}
{\partial}_{\nu} F^{\nu \mu} =  J^{\mu},
\label{eq: dvadva}
\end{equation}
where the total matter current
\[
J^{\mu} =\sum_{k=1}^2 e_k \bar{\psi}_k {\gamma}^{\mu} {\psi}_k
\]
is conserved, ${\partial}_{\mu} J^{\mu} =0$.

In the Coulomb gauge $A_1(x,t)=0$, the equations (\ref{eq: dvadva})
take the form
\begin{eqnarray*}
{\partial}_1^2 A_0 & = & - J^0,\\
{\partial}_1 {\partial}_0 A_0 & = & J^1.
\end{eqnarray*}
These two equations reduce in fact to each other because of
the total current conservation  and are solved by
\begin{equation}
A_0(x,t) = - \int_{- \infty}^{\infty} dy D(x,y) J^0(y,t),
\label{eq: dvatri}
\end{equation}
where the Green's function is
\[
D(x,y) = \frac{1}{2} |x-y|.
\]

In contrast with the situation on the circle,
the electromagnetic field on the line has not a global
physical degree of freedom and can be therefore eliminated
from the action completely.
If we insert (\ref{eq: dvatri}) into Eq.(\ref{eq: dvaodin}), we obtain
the action in the Coulomb gauge as
\[
W[{\psi},A] = \int_{- \infty}^{\infty} dt \int_{- \infty}^{\infty}
dx \sum_{k=1}^2 \overline{\psi}_k ({\gamma}^{\mu} 
i {\hbar}c {\partial}_{\mu}
- m_k c^2) {\psi}_k +
\]
\begin{equation}
+ \frac{1}{2} \int_{- \infty}^{\infty} dt \int_{- \infty}^{\infty}
dx \int_{- \infty}^{\infty} dy J^0(x,t) D(x,y) J^0(y,t).
\label{eq: dvacet}
\end{equation}

We could vary the action (\ref{eq: dvacet}) with respect to individual
fields ${\psi}_1$ and ${\psi}_2$ separately. This results in non-linear
coupled Hartree-type equations for these fields. Instead, we use a
relativistic configuration space formalism \cite{bar90,aob90}
to take into account the long-range quantum correlations. We define
a composite field ${\Phi}$ by
\[
{\Phi}(x_1,t|x_2,t) \equiv {\psi}_1(x_1,t) \otimes   {\psi}_2(x_2,t)
\]
which is a four-component spinor field. The configuration space $(x_1,x_2)$
is two-dimensional Euclidean space ${\rm R}^2$.

We can rewrite our action (\ref{eq: dvacet}) entirely in terms of the
composite field ${\Phi}$. The resultant action is  \cite{barsa94}
\[
{\rm W}[\Phi, A] = \int_{- \infty}^{\infty} dt \int_{- \infty}^{\infty} dx_1
\int_{- \infty}^{\infty} dx_2 \bar{\Phi}(x_1,t|x_2,t)
\{ (c {\gamma}^{\mu} p_{(1),\mu} -m_1 c^2) \otimes {\gamma}^0 +
{\gamma}^0 \otimes (c {\gamma}^{\mu} p_{(2),\mu} -m_2 c^2)
\]
\begin{equation}
+ \frac{1}{2} ({\gamma}^0 \otimes {\gamma}^0)
(e_1 {\phi}_{(1)}^{\rm self} + e_2 {\phi}_{(2)}^{\rm self})
+e_1e_2 ({\gamma}^0 \otimes {\gamma}^0) D(x_1,x_2) \} 
{\Phi}(x_1,t|x_2,t) ,
\label{eq: dvapet}
\end{equation}
where
\[
p_{(i),\mu} \equiv  i{\hbar} \frac{\partial}{\partial x_{i}^{\mu}},
\]
and
\begin{eqnarray*}
{\phi}_{(1)}^{\rm self}(x,t) & = &
e_1 \int_{- \infty}^{\infty} dy \int_{- \infty}^{\infty} dz 
D(x,z) \bar{\Phi}(z,t|y,t) ({\gamma}^0 \otimes {\gamma}^0)
{\Phi}(z,t|y,t), \\
{\phi}_{(2)}^{\rm self}(x,t) & = &
e_2 \int_{- \infty}^{\infty} dy \int_{- \infty}^{\infty} dz 
D(x,y)
\bar{\Phi}(z,t|y,t) ({\gamma}^0 \otimes {\gamma}^0)
{\Phi}(z,t|y,t),\\
\end{eqnarray*}
the self-potentials ${\phi}_{(k)}^{\rm self}$ being non-linear integral
expressions. The spin matrices are written here in the form of tensor
products $ \otimes $ , the first factor always referring to the spin
space of particle 1, the second to particle 2.

Now we require the action (\ref{eq: dvapet}) to be stationary
not with respect to the variation of the individual fields
but with respect to the composite field only. This leads to the
following two-body wave equation 
\[
\{ ({\gamma}^{\mu} {\pi}_{(1),\mu} - m_1 c) \otimes {\gamma}^0 +
{\gamma}^0 \otimes ({\gamma}^{\mu} {\pi}_{(2),\mu} -m_2 c)
\]
\begin{equation}
+ \frac{e_1e_2}{c} ({\gamma}^0 \otimes {\gamma}^0) D(x_1,x_2) \} 
{\Phi}(x_1,t|x_2,t) =0,
\label{eq: dvashest}
\end{equation}
where the generalized (kinetic) momenta ${\pi}_{(i),\mu}$
are given by
\[
{\pi}_{(i),\mu} = p_{(i),\mu} + \frac{e_i}{c} A_{(i),\mu}^{\rm self}
\]
with
\[
A_{(1),0}^{\rm self} \equiv {\phi}_{(1)}^{\rm self} ,
\hspace{5 mm} A_{(2),0}^{\rm self} \equiv {\phi}_{(2)}^{\rm self},
\]
and
\[
A_{(1),1}^{\rm self} = A_{(2),1}^{\rm self}=0.
\]

In the center of mass and relative coordinates
\begin{eqnarray*}
{\Pi} = {\pi}_{(1)} + {\pi}_{(2)} & , & {\pi} = {\pi}_{(1)} - {\pi}_{(2)}, \\
P = p_{(1)} + p_{(2)} & , & p = p_{(1)} - p_{(2)},\\
x_{+} = x_1 + x_2 & , & x_{-} = x_1 - x_2, \\
\end{eqnarray*}
the function $D(x_1,x_2)$ becomes
\[
D(x_1,x_2) =  D_{-}(x_{-}) = \frac{1}{2} |x_{-}|,
\]
i.e. depends only on the relative coordinate $x_{-}$ and is
symmetric.

The two-body equation , first without the self-field terms, 
takes the form
\[
\left[ {\Gamma}^{\mu} P_{\mu} + k^{\mu} p_{\mu} + \frac{e_1e_2}{c}
({\gamma}^0 \otimes {\gamma}^0) D_{-}(x_{-}) -
m_1 c I \otimes {\gamma}^0 - m_2 c {\gamma}^0 \otimes I \right]
{\Phi}(x_{-},t|x_{+},t) =0,
\]
where
\[
{\Gamma}^{\mu} \equiv \frac{1}{2} ( {\gamma}^{\mu} \otimes
{\gamma}^0 + {\gamma}^0 \otimes {\gamma}^{\mu} ),
\]
\[
k^{\mu} \equiv \frac{1}{2} ( {\gamma}^{\mu} \otimes {\gamma}^0
- {\gamma}^0 \otimes {\gamma}^{\mu} ),
\]
and $I$ is identity matrix.
We see that $k^0$ vanishes which means that the relative energy
$p_0$ drops out of the equation automatically and we get
\begin{equation}
\left[ {\Gamma}^0 P^0 - {\Gamma}^1 P^1 - k^1 p^1 + \frac{e_1e_2}{c}
({\gamma}^0 \otimes {\gamma}^0) D_{-}(x_{-}) -
m_1 c I \otimes {\gamma}^0 - m_2 c {\gamma}^0 \otimes I \right]
{\Phi}(x_{-},t|x_{+},t) =0.
\label{eq: dvasem}
\end{equation}
Thus we have only one time variable conjugate to the center
of mass energy $\frac{1}{c}P^0$, one degree of freedom for the center
of mass momentum $P^1$ and one degree of freedom for the
relative momentum $p^1$. Since $P^0$ is the "Hamiltonian"
of the system, by multiplying (\ref{eq: dvasem}) by
${\Gamma}_0^{-1}$ we obtain the Hamiltonian form of the
two-body equation
\begin{equation}
P_0 {\Phi} = \left( {\alpha}_{+} P^1 + {\alpha}_{-} p^1 
- \frac{e_1e_2}{c} D_{-} + {\beta}_1 m_1 c +
{\beta}_2 m_2 c \right) {\Phi},
\label{eq: dvavosem}
\end{equation}
with
\[
{\alpha}_{\pm} \equiv \frac{1}{2} ({\alpha}_1 \pm {\alpha}_2),
\hspace{5 mm}
{\alpha}_1 \equiv {\gamma}^5 \otimes I,
\hspace{5 mm}
{\alpha}_2 \equiv I \otimes {\gamma}^5,
\hspace{5 mm}
{\beta}_1 \equiv {\gamma}^0 \otimes I,
\hspace{5 mm}
{\beta}_2 \equiv I \otimes {\gamma}^0,
\]
and the relative and center of mass terms  in the Hamiltonian
$P_0$ being additive:
\[
P_0 = H_{\rm c.m.} + H_{\rm rel},
\]
\[
H_{\rm c.m.} \equiv {\alpha}_{+} P^1 ,
\]
\[
H_{\rm rel} \equiv {\alpha}_{-} p^1
- \frac{e_1e_2}{c} D_{-} + {\beta}_1 m_1 c + {\beta}_2 m_2 c.
\]
Eq.(\ref{eq: dvavosem}) has the form of a generalized
Dirac
equation, now a four-component wave equation.

With the self-potential terms the Hamiltonian form of the
two-body equation becomes
\[
P_0 {\Phi} = \left( {\alpha}_{+} P^1 + {\alpha}_{-} p^1 
- \frac{1}{c} {\phi}_{-} - \frac{e_{1}}{c} {\phi}_{(1)}^{\rm self} -
\frac{e_{2}}{c} {\phi}_{(2)}^{\rm self} + {\beta}_1 m_1 c +
{\beta}_2 m_2 c \right) {\Phi},
\]
where
\[
{\phi}_{-} = e_{1} e_{2} D_{-}.
\]
The self-potentials break in general the above mentioned additivity
of the center of mass and relative parts of $P_0$.

\renewcommand{\ren}{\renewcommand{\theequation}{3.\arabic{equation}}}
\renewcommand{\add}{\addtocounter{equation}{-1}}
\renewcommand{\new}[1]{\renewcommand{\theequation}{3.\arabic{equation}#1}}
\newcommand{\set}{\setcounter{equation}{0}}

\set

\begin{center}

\section{SPECTRUM}
\label{sec: anal}

\end{center}

Let us find the eigenfunctions and the spectrum of $P_0$. 
The equation for the eigenfunctions is
\ren
\begin{equation}
({\alpha}_{+} P^1 + {\alpha}_{-} p^1
+ {\beta}_1 m_1 c + {\beta}_2 m_2 c ) {\Phi} =
\frac{1}{c} (E+\tilde{\phi}) {\Phi},
\label{eq: triodin}
\end{equation}
where
\[
P^1 = 2i {\hbar} \frac{\partial}{\partial x_{+}},
\hspace{5 mm}
p^1 = 2i {\hbar} \frac{\partial}{\partial x_{-}},
\]
and
\[
\tilde{\phi} \equiv {\phi}_{-} + e_{1} {\phi}_{(1)}^{\rm self} +
e_2 {\phi}_{(2)}^{\rm self}.
\]
If we denote the components of the composite field ${\Phi}$ as
\begin{eqnarray*}
{\Phi}^{11} \equiv {\eta}_1 & , & {\Phi}^{12} \equiv {\eta}_2,\\
{\Phi}^{21} \equiv {\eta}_3 & , & {\Phi}^{22} \equiv {\eta}_4,\\
\end{eqnarray*}
then (\ref{eq: triodin}) reduces to the system of four equations:
\[
2i {\hbar} \frac{\partial}{\partial x_{+}} {\eta}_1 - 
\frac{1}{c} ( \tilde{\phi} +E) {\eta}_1
= -m_1 c {\eta}_3 - m_2 c {\eta}_2, 
\]
\new{a}
\begin{equation}
2i {\hbar} \frac{\partial}{\partial x_{+}} {\eta}_4 + 
\frac{1}{c} ( \tilde{\phi} +E) {\eta}_4
= m_1 c {\eta}_2 + m_2 c {\eta}_3, 
\end{equation}
\[
2i {\hbar} \frac{\partial}{\partial x_{-}} {\eta}_2 - 
\frac{1}{c} ( \tilde{\phi} +E) {\eta}_2
= -m_1 c {\eta}_4 - m_2 c {\eta}_1,
\]
\add
\new{b}
\begin{equation}
2i {\hbar} \frac{\partial}{\partial x_{-}} {\eta}_3 + 
\frac{1}{c} ( \tilde{\phi} +E) {\eta}_3
= m_1 c {\eta}_1 + m_2 c {\eta}_4. 
\end{equation}

We see from these equations that
\new{a}
\begin{equation}
{\eta}_1^{\star}(E,e_1,e_2) = {\eta}_4(E,e_1,e_2),
\end{equation}
\add
\new{b}
\begin{equation}
{\eta}_2^{\star}(E,e_1,e_2) = {\eta}_3(E,e_1,e_2),
\end{equation}
i.e. only half of all solutions of Eqs.(3.2a) -
(3.2b) are independent and correspond to physical
particles.

It is more convenient to introduce the combinations
\begin{eqnarray*}
{\eta}_{\pm} & \equiv & {\eta}_2 \pm {\eta}_3, \\
{\chi}_{\pm} & \equiv & {\eta}_1 \pm {\eta}_4,
\end{eqnarray*}
and use them instead of the original componenets
${\eta}_{i}$ $(i=\overline{1,4})$. The system of
equations (3.2a-b) takes the form
\[
2i {\hbar} \frac{\partial}{\partial x_{+}} {\chi}_{+} 
- \frac{1}{c} f {\chi}_{-}=
({\Delta}m) c {\eta}_{-},
\]
\ren
\new{a}
\begin{equation}
2i {\hbar} \frac{\partial}{\partial x_{+}} {\chi}_{-} 
- \frac{1}{c} f {\chi}_{+}=
- M c {\eta}_{+},
\end{equation}
\[
2i {\hbar} \frac{\partial}{\partial x_{-}} {\eta}_{+} 
- \frac{1}{c} f {\eta}_{-}=
({\Delta}m) c {\chi}_{-},
\]
\add
\new{b}
\begin{equation}
2i {\hbar} \frac{\partial}{\partial x_{-}} {\eta}_{-} 
- \frac{1}{c} f {\eta}_{+}=
- M c {\chi}_{+},
\end{equation}
where $M \equiv m_1+m_2$, ${\Delta}m \equiv m_1 - m_2$, and
$f = \tilde{\phi} + E$. 

Without loss of generality, we can
take masses equal to each other, $m_1 = m_2 \equiv m$.
Then the eigenfunctions
${\chi}_{-}$ and ${\eta}_{-}$ are determined by ${\chi}_{+}$
and ${\eta}_{+}$, respectively,
\begin{eqnarray*}
{\chi}_{-} & = & \frac{1}{f} 2i {\hbar}c 
\frac{\partial}{\partial x_{+}} {\chi}_{+},\\
{\eta}_{-} & = & \frac{1}{f} 2i {\hbar}c 
\frac{\partial}{\partial x_{-}} {\eta}_{+},
\end{eqnarray*}
while ${\chi}_{+}$ and ${\eta}_{+}$ are related by
\new{a}
\begin{equation}
T_{-} {\eta}_{+} = - 2mc {\chi}_{+},
\end{equation}
\add
\new{b}
\begin{equation}
T_{+} {\chi}_{+} = - 2mc {\eta}_{+},
\end{equation}
where
\ren
\begin{equation}
T_{\pm} \equiv -4 {\hbar}^2c \frac{\partial}{\partial x_{\pm}}
\frac{1}{f} \frac{\partial}{\partial x_{\pm}} - \frac{1}{c} f.
\label{eq: trishest}
\end{equation}

Acting on (3.5a) by $T_{-}$ and on (3.5b) by $T_{+}$,
we easily decouple ${\chi}_{+}$ and ${\eta}_{+}$ and
rewrite (3.5a-b) equivalently as
\new{a}
\begin{equation}
T_{+}T_{-} {\eta}_{+} = 4m^2c^2 {\eta}_{+},
\end{equation}
\add
\new{b}
\begin{equation}
T_{-}T_{+} {\chi}_{+} = 4m^2c^2 {\chi}_{+}.
\end{equation}

In what follows we assume for the eigenfunctions the ansatz
\begin{eqnarray*}
{\eta}_{\pm}(x_{-},x_{+}) & = & 
\exp\left( \frac{i}{\hbar} P_{cm} x_{+} \right) {\eta}_{\pm}(x_{-}),\\
{\chi}_{\pm}(x_{-},x_{+}) & = & 
\exp\left( \frac{i}{\hbar} P_{cm}x_{+} \right) {\chi}_{\pm}(x_{-}),
\end{eqnarray*}
i.e. separate their $x_{+}$-dependent part from the
$x_{-}$-dependent one. Here $P_{cm}$ is a momentum
conjugate to the  center of mass coordinate $x_{+}$, so we
can call it "the center of mass motion momentum".

\begin{center}

\subsection{Free motion}

\end{center}

If we neglect both mutual and self-interactions, i.e.
put $\tilde{\phi}=0$, then we get a system of two
"free" particles. The relations (3.3a-b) become
\begin{eqnarray*}
{\eta}_1(-E) & = & {\eta}_4(E), \\
{\eta}_2(-E) & = & - {\eta}_3(E), \\
\end{eqnarray*}
i.e. the negative energy solutions of ${\eta}_1$ and
${\eta}_2$ coincide correspondingly with the positive
energy solutions of ${\eta}_4$ and ${\eta}_3$. Therefore
we may consider either positive and negative energy
solutions of ${\eta}_1$ and ${\eta}_2$ or only positive
energy solutions of all four equations (3.2a-b) as
physical particles.

For $\tilde{\phi}=0$, the operators $T_{-}$ and $T_{+}$
commute, so Eqs.(3.7a) and (3.7b) coincide with each other.
Their solution is
\begin{eqnarray*}
{\eta}_{+}(x_{-}) & = & \sin\left(\frac{1}{\hbar}{\kappa}x_{-}\right),\\
{\chi}_{+}(x_{-}) & = & \frac{2mc^2E}{E^2-4c^2P_{cm}^2}
\sin\left(\frac{1}{\hbar} {\kappa} x_{-} \right),
\end{eqnarray*}
where
\[
{\kappa} \equiv \frac{E}{2c} \sqrt{\frac{E^2-4c^2P_{cm}^2
-4m^2c^4}{E^2-4c^2P_{cm}^2}}.
\]
For the other two components, we have
\begin{eqnarray*}
{\eta}_{-}(x_{-}) & = & i \frac{2c{\kappa}}{E} 
\cos\left( \frac{1}{\hbar} {\kappa} x_{-} \right), \\
{\chi}_{-}(x_{-}) & = & - \frac{4mc^3P_{cm}}{E^2 - 4c^2P_{cm}^2}
\sin\left( \frac{1}{\hbar} {\kappa} x_{-} \right).
\end{eqnarray*}
This solution exists for all energies for which
\[
|E| \ge 2mc^2 \sqrt{1+ \frac{P_{cm}^2}{m^2c^2}},
\]
and there is a "forbidden" band in between (see Fig. 1a).

\begin{center}

\subsection{Coulomb interaction}

\end{center}

Let us now consider our two-body system in the presence of the
Coulomb interaction and neglect only self-field terms. The
relations (3.3a-b) take the form
\begin{eqnarray*}
{\eta}_1(-E , -e_1e_2) & = & {\eta}_4(E , e_1e_2), \\
{\eta}_2(-E , -e_1e_2) & = & - {\eta}_3(E , e_1e_2),
\end{eqnarray*}
i.e. the negative energy solutions of ${\eta}_1$ and ${\eta}_2$
coincide correspondingly with the positive energy solutions
of ${\eta}_4$ and ${\eta}_3$ of opposite sign of $e_1e_2$.

For $\tilde{\phi} = {\phi}_{-}$, the equation (3.7a) for
${\eta}_{+}$ reduces to the second order differential
equation
\ren
\begin{equation}
\frac{{\partial}^2{\eta}_{+}}{{\partial} x_{-}^2} -
\frac{1}{f} \left( \frac{{\partial}f}{{\partial}x_{-}} \right)
\frac{{\partial}{\eta}_{+}}{{\partial}x_{-}}
+ \frac{1}{4{\hbar}^2c^2} f^2{\eta}_{+} =
\frac{m^2c^2}{{\hbar}^2} \frac{f^2}{f^2-4c^2P_{cm}^2} {\eta}_{+}.
\label{eq: trivosem}
\end{equation}
If we make in (\ref{eq: trivosem}) the substitution
\[
{\eta}_{+}(x_{-}) = \sqrt{f} \cdot \sigma(x_{-}),
\]
then we find for $\sigma$ the following Schr\"odinger type
equation
\ren
\begin{equation}
- \frac{d^2{\sigma}}{dx_{-}^2} + V(x_{-}) \sigma = K \sigma
\label{eq: tridevet}
\end{equation}
with the "potential"
\[
V(x_{-}) \equiv - \frac{1}{2f} \frac{d^2f}{dx_{-}^2} +
\frac{3}{4} \left(\frac{1}{f} \frac{df}{dx_{-}}\right)^2
- \frac{1}{4{\hbar}^2c^2} f^2 + \frac{4m^2c^4P_{cm}^2}
{{\hbar}^2(f^2 - 4c^2P_{cm}^2)}
\]
and the "energy"
\[
K \equiv -\frac{m^2c^2}{{\hbar}^2}.
\]
The last term in the potential represents the center of mass
motion contribution which vanishes for $P_{cm}=0$ as well as
for all values of $P_{cm}$ in the massless case.

The explicit form of the potential for $P_{cm}=0$ is
\[
V(x_{-}) = - \frac{1}{s} \delta(x_{-}) + \tilde{V}(x_{-}),
\]
where $s \equiv \frac{2E}{e_1e_2}$, and
\ren
\begin{equation}
\tilde{V}(x_{-}) = \frac{3}{4} \frac{1}{(|x_{-}|+s)^2}
- \frac{1}{16{\hbar}^2c^2} (e_1e_2)^2 (|x_{-}|+s)^2.
\label{eq: trideset}
\end{equation}
The potential $V(x_{-})$ has several pecularities. First,
it contains a $\delta$-functional part with coefficient
$(- 1/s)$ , positive ( for $e_1e_2>0$ and $E<0$ or $e_1e_2<0$
and $E>0$ ) or negative ( for $e_1e_2>0$ and $E>0$ or
$e_1e_2<0$ and $E<0$ ). The form of $V(x_{-})$ for different
signs of the $\delta$-function coefficient is shown in
Figures 2 and 3. Secondly, its regular part $\tilde{V}(x_{-})$
includes the inverted $x^2$-potential (the last term in
Eq.(\ref{eq: trideset}) ). Such kind of potentials is known
to appear in barrier penetration problems, splitting in
double wells, and tunneling out of traps \cite{ford59}-
\cite{cald83}. The
systematic study of the inverted oscillator 
is given in \cite{bart86}. The presence of the inverted
$x^2$-potential makes $V(x_{-})$ nonvanishing at $|x_{-}| 
\to \infty$ and indicates that the particles are not
confined in a stable system.

The regular part $\tilde{V}(x_{-})$ is symmetric with
respect to $x_{-}$, $\tilde{V}(-x_{-}) = \tilde{V}(x_{-})$.
For very small non-zero $x_{-}$, $|x_{-}| \ll |s|$ ,
$\tilde{V}(x_{-})$ is approximately linear
\[
\tilde{V}(x_{-}) \approx \tilde{V}_0 -
\left[ \frac{3}{2} \frac{1}{s^2} + \frac{1}{8{\hbar}^2c^2}
(e_1e_2)^2s \right] |x_{-}|,
\]
where
\[
\tilde{V}_0 \equiv \tilde{V}(x_{-}=0) = \frac{3}{4} \frac{1}{s^2}
- (\frac{e_1e_2}{4{\hbar}c})^2 s^2.
\]
The value $\tilde{V}_0$ is positive for $E^2<\frac{\sqrt{3}}{2}
{\hbar}c |e_1e_2|$ and negative for $E^2>\frac{\sqrt{3}}{2}
{\hbar}c |e_1e_2|$.

While ${\sigma}(x_{-})$ is continuous for all $x_{-}$, its
first derivative $d{\sigma}/dx_{-}$ changes discontinuously
at the point $x_{-}=0$. This is because of the $\delta$-
functional potential in the Schr\"odinger equation for
$\sigma$. If we integrate both parts of (\ref{eq: tridevet})
over infinitely small interval $(-{\varepsilon},{\varepsilon})$,
${\varepsilon} \ll 1$, and then take the limit ${\varepsilon}
\to 0_{+}$, we get the matching condition
\ren
\begin{equation}
\frac{d{\sigma}}{dx_{-}}(+0) - \frac{d{\sigma}}{dx_{-}}(-0)
= - \frac{1}{s} {\sigma}(0).
\label{eq: triodinodin}
\end{equation}

The Schr\"odinger equation (\ref{eq: tridevet}) taken without
the center of mass motion contribution to the potential can be
solved exactly. With the substitution
\[
{\sigma}(x_{-}) = z^{3/4} \exp\left(- i \frac{z}{2{\hbar}c} 
\right) u(z),
\]
where $z \equiv \frac{1}{4} e_1e_2 (|x_{-}|+s)^2$, and away from
the origin $(x_{-} \neq 0$ or $z \neq z_0 \equiv \frac{1}{4}
e_1e_2 s^2 = {E^2}/{e_1e_2} )$ the equation becomes 
\[
z \frac{d^2u}{dz^2} + (2-i\frac{z}{{\hbar}c}) \frac{du}{dz}
- (\frac{i}{{\hbar}c} - \frac{K}{e_1e_2})u =0.
\]
The first independent solution of this equation is
\[
u_1 = F(1+i{\beta},2; \frac{iz}{{\hbar}c}),
\hspace{1 cm}
\beta \equiv \frac{K{\hbar}c}{e_1e_2},
\]
i.e. the confluent hypergeometric function. The integral
representation for $u_1$ is \cite{dau74}
\[
u_1 = 2\exp\left( i\frac{z}{2{\hbar}c} \right) 
{\rm Re}\left[ \frac{1}{{\Gamma}
(1+i{\beta})} \exp\left( i\frac{z}{2{\hbar}c} \right) 
(\frac{iz}{{\hbar}c})
^{-1+i{\beta}} G(1-i{\beta}, -i{\beta}; 
\frac{iz}{{\hbar}c} ) \right] .
\]
The definition and some useful formulae for the function $G$
are given in Appendix.
The asymptotic behaviour of the first solution is
\[
{\sigma}_1(|z| \to \infty) \approx \frac{2{\hbar}ce^{-\frac{\pi}{2}
{\beta}}}{|{\Gamma}(1+i{\beta})|} z^{-1/4} \sin\left(\frac{z}{2{\hbar}c}
+ {\beta}{\rm ln}\frac{z}{{\hbar}c} + \delta\right),
\]
\[
\delta \equiv {\rm arg}{\Gamma}(1-i{\beta}).
\]

The second independent solution is
\[
u_2 = -2\exp\left( i\frac{z}{2{\hbar}c} \right) 
{\rm Im}\left[  \frac{1}{{\Gamma}
(1+i{\beta})} \exp\left( i\frac{z}{2{\hbar}c} \right) 
(\frac{iz}{{\hbar}c})
^{-1+i{\beta}} G(1-i{\beta},-i{\beta};
\frac{iz}{{\hbar}c}) \right] .
\]
Its asymptotic behaviour is
\[
{\sigma}_2 (|z| \to \infty) \approx \frac{ 2{\hbar}c
e^{-\frac{\pi}{2}{\beta}} } {| {\Gamma}(1+i{\beta}) |}
z^{-1/4} \cos\left(\frac{z}{2{\hbar}c} 
+ \beta {\rm ln}\frac{z}{{\hbar}c}
+ \delta \right).
\]

If we write the total solution
\[
\sigma = A \sigma_1 + B \sigma_2,
\]
where $A$ and $B$ are arbitrary constants, and use the matching
condition (\ref{eq: triodinodin}) which in terms of $z$
becomes
\[
\frac{{\sigma}^{\prime}(z_0)}{{\sigma}(z_0)} =
- \frac{1}{4} \frac{1}{z_0},
\]
then we get the following relation between $A$ and $B$ :
\[
\frac{A}{B} = - \frac{4z_0{\sigma}_2^{\prime}(z_0) +
{\sigma}_2(z_0)}{4z_0{\sigma}_1^{\prime}(z_0) +
{\sigma}_1(z_0)},
\]
the prime indicating the derivation with respect to $z$.

Asymptotically the total solution behaves as
\[
{\sigma}(|z| \to \infty) \approx \frac{{\hbar}ce^{-\frac{\pi}
{2}{\beta}}}{|{\Gamma}(1+i{\beta})|} z^{-1/4}
\{ (B-iA) \exp\left(i \frac{z}{2{\hbar}c} + i{\beta}
{\rm ln} \frac{z}{{\hbar}c} + i\delta \right)  
\]
\[ 
+ (B+iA) \exp\left(-i \frac{z}{2{\hbar}c} -i{\beta}
{\rm ln}\frac{z}{{\hbar}c} -i\delta \right) \}.
\]
Regardless of values of the constants $A$ and $B$,
$\sigma$ does not vanish at $|z| \to \infty$ and can not
represent bound states. For $B+iA=0$, the solution
behaves at infinity as a diverging wave. Such behaviour
is specific for metastable states.
The condition determining the metastable (or quasi-discrete)
energy levels is then
\ren
\begin{equation}
\frac{{\sigma}^{\prime}_1(z_0) - i {\sigma}_2^{\prime}(z_0)}
{{\sigma}_1(z_0) -i {\sigma}_2(z_0)} =
- \frac{1}{4z_0}.
\label{eq: triodindva}
\end{equation}
With the expressions for ${\sigma}_1$ and ${\sigma}_2$, we
rewrite (\ref{eq: triodindva}) as
\[
\frac{G^{\prime}(1-i{\beta},-i{\beta};\frac{iz_0}{{\hbar}c})}
{G(1-i{\beta},-i{\beta};\frac{iz_0}{{\hbar}c})} =
-i(\frac{1}{2{\hbar}c} + \frac{\beta}{z_0}),
\]
and then, using the relation between $G$ and its derivative
(see Appendix), finally come to
\ren
\begin{equation}
i{\hbar}c \frac{{\beta}(1-i{\beta})}{z_0^2} \cdot
\frac{G(2-i{\beta},1-i{\beta};\frac{iz_0}{{\hbar}c})}
{G(1-i{\beta},-i{\beta};\frac{iz_0}{{\hbar}c})} =
\frac{1}{2{\hbar}c} + \frac{\beta}{z_0}.
\label{eq: triodintri}
\end{equation}

Metastable states are described by complex values of energy
\[
E = E_0 - i \frac{\Gamma}{2},
\]
where $E_0$ is the metastable level energy, while ${\Gamma}$
is its width. For $z_0$ we get
\[
z_0 = \frac{1}{e_1e_2} (E_0^2 - \frac{{\Gamma}^2}{4}) -
i \frac{E_0 {\Gamma}}{e_1e_2}.
\]

We can solve Eq.(\ref{eq: triodintri}) approximately for large
values of $m$, $m^2 \gg {\hbar}|e_1e_2|/c^3 $. In this
approximation, $|{\beta}| \gg 1$, so
\[
\frac{G(2-i{\beta},1-i{\beta};\frac{iz_0}{{\hbar}c})}
{G(1-i{\beta},-i{\beta};\frac{iz_0}{{\hbar}c})}
\approx 1.
\]
Eq.(\ref{eq: triodintri}) takes the form
\[
i{\beta}(1-i{\beta}) = \frac{1}{2} (\frac{z_0}{{\hbar}c})^2
+ {\beta} (\frac{z_0}{{\hbar}c}),
\]
which is solved by
\[
E_0^{(1)} \approx \pm mc^2 \sqrt{1+\sqrt{3}}  \cdot
\left[1 + \frac{1}{12} (1- \frac{5\sqrt{3}}{12}) \frac{(e_1e_2{\hbar})^2}
{m^4c^6}\right],
\]
\[
{\Gamma}^{(1)} \approx \frac{|e_1e_2|{\hbar}}{\sqrt{3}mc} \cdot
\frac{1}{\sqrt{1+\sqrt{3}}},
\]
and
\[
E_0^{(2)} \approx \pm \frac{e_1e_2{\hbar}}{\sqrt{6}mc}
\cdot \frac{1}{(1+\sqrt{3})^{3/2}},
\]
\[
{\Gamma}^{(2)} \approx \sqrt{2} mc^2 \cdot (1+\sqrt{3})^{3/2}.
\]
There are therefore four metastable energy levels in the band
between the positive and negative energy continuums, the
first two at the energies $E_0^{(1)} \approx \pm mc^2
\sqrt{1+\sqrt{3}} = \pm 1.65 \times mc^2$ and the other two
at the energies close to zero, $E_0^{(2)} \approx 0_{\pm}$
(see Fig.1b). For infinitely large values of mass,
$m \to \infty$, the first two metastable energy levels
turn into stable ones, ${\Gamma}^{(1)} \to 0$, while
the second two disappear, ${\Gamma}^{(2)} \to \infty$.

For $e_1e_2 >0$, the positive energy metastable levels
correspond to the relative motion in the potential
$V(x_{-})$ with $s>0$, while the negative energy ones
in the potential with $s<0$. For $e_1e_2<0$, the
metastable energy levels are positive for $s<0$ and
negative for $s>0$.

In the massless case $({\beta}=0)$, the solutions
${\sigma}_1$ and ${\sigma}_2$ become trigonometric functions
\begin{eqnarray*}
{\sigma}_1 & = & 2{\hbar}c z^{-1/4} \sin\left(\frac{z}
{2{\hbar}c}\right),\\
{\sigma}_2 & = & 2{\hbar}c z^{-1/4} \cos\left(\frac{z}
{2{\hbar}c}\right).
\end{eqnarray*}
The analogue of the condition (\ref{eq: triodindva}) is
\[
{\rm tg}\left(\frac{z_0}{2{\hbar}c}\right) = - \frac{i}{2}.
\]
This condition does not fix $E_0$, while for the level
width we get ${\Gamma} = \infty$. This means that in the
massless case there are neither discrete nor quasi-discrete
energy levels and the energy spectrum is continuous.

The spectrum for ${\chi}_{+}$ can be derived analogously.
It can be shown that the equation (3.7b) for ${\chi}_{+}$
also reduces to the Schr\"odinger type equation with the
potential
\[
U(x_{-}) = V(x_{-}) + \left[ \frac{1}{f} \frac{{\partial}^2f}
{{\partial}x_{-}^2} - \left(\frac{1}{f} \frac{{\partial}f}
{{\partial}x_{-}}\right)^2 \right] \cdot \frac{4c^2P_{cm}^2 + f^2}
{4c^2P_{cm} - f^2} - \frac{8}{f^2} \cdot \frac{c^2P_{cm}^2}
{4c^2P_{cm}^2 - f^2} \left(\frac{{\partial}f}
{{\partial}x_{-}}\right)^2.
\]
For $P_{cm}=0$, the explicit form of $U(x_{-})$ is
\[
U(x_{-}) = -\frac{3}{s} \delta(x_{-}) + \tilde{U}(x_{-}),
\]
\[
\tilde{U}(x_{-}) = \frac{7}{4} \frac{1}{(|x_{-}|+s)^2}
- \frac{1}{16} (\frac{e_1e_2}{{\hbar}c})^2
(|x_{-}|+s)^2.
\]
Acting along similar lines as above, we get the following
metastable spectrum equation
\ren
\begin{equation}
\frac{G^{\prime}(\frac{1}{2} + \frac{1}{\sqrt{2}} -i{\beta},
\frac{1}{2} - \frac{1}{\sqrt{2}} -i{\beta}; \frac{iz_0}{{\hbar}c})}
{G(\frac{1}{2} + \frac{1}{\sqrt{2}} -i{\beta},
\frac{1}{2} - \frac{1}{\sqrt{2}} -i{\beta}; \frac{iz_0}{{\hbar}c} )}
= - \frac{i}{2{\hbar}c} - (\frac{1}{2} +i{\beta})
\frac{1}{z_0}.
\label{eq: triodincet}
\end{equation}
For large values of $m$, this equation is solved by
\[
E_0^{(1)} \approx \pm mc^2 \sqrt{1+\sqrt{3}} \cdot 
\left[1 + \frac{1}{48} (1- \frac{1}{2\sqrt{3}}) \frac{(e_1e_2{\hbar})^2}
{m^4c^6}\right],
\]
\[
{\Gamma}^{(1)} \approx \frac{{\hbar}|e_1e_2|}{2\sqrt{3}mc}
\cdot \sqrt{1+\sqrt{3}},
\]
and
\[ 
E_0^{(2)} \approx \pm \frac{e_1e_2{\hbar}}{2\sqrt{6}mc}
\cdot \frac{1}{\sqrt{1+\sqrt{3}}},
\]
\[
{\Gamma}^{(2)} \approx 2\sqrt{2} mc^2 \cdot 
\frac{1}{\sqrt{1+\sqrt{3}}}.
\]
The structure of the spectrum for ${\chi}_{+}$ 
is the same as in the case of
${\eta}_{+}$. For infinitely large values of $m$, the spectrums
for ${\eta}_{+}$ and ${\chi}_{+}$ coincide exactly, while
for large and finite values of $m$, the corrections of the order
$(1/{\beta})$ and $(1/{{\beta}^2})$ are different.

\begin{center}

\subsection{Self-interaction}

\end{center}

The self-interaction makes the spectrum problem essentially
more complicated. Let us give here a few comments concerning
the effects of the self-field terms.

With the self-potentials ${\phi}_{(1)}^{\rm self}$ ,
${\phi}_{(2)}^{\rm self}$, the function $f$ depends on both
coordinates $x_{-}$ and $x_{+}$, and the operators $T_{\pm}$
acquire additional terms including the partial derivative
$({\partial}f/{\partial}x_{+})$. Moreover, we can not
assume, as before in the study of the Coulomb interaction,
that the center of mass motion is free, with a momentum
$P_{cm}$. This results in infinitely large values of the
self-potentials.

Indeed, in terms of the components ${\eta}_i$ $(i=\overline{1,4})$
the self-potentials take the form
\begin{eqnarray*}
{\phi}_{(1)}^{\rm self}(x) & = & \frac{e_1}{2} \int_{-\infty}
^{\infty} dy_{-} \int_{-\infty}^{\infty} dy_{+} D\left(x,\frac{1}{2}
(y_{+} + y_{-}) \right) \sum_{i=1}^{4} {\eta}_i^{\star}(y_{-},y_{+})
{\eta}_i(y_{-},y_{+}), \\
{\phi}_{(2)}^{\rm self}(x) & = & \frac{e_2}{2} \int_{- \infty}
^{\infty} dy_{-} \int_{-\infty}^{\infty} dy_{+} D\left(x,\frac{1}{2}
(y_{+} - y_{-}) \right) \sum_{i=1}^{4} {\eta}_i^{\star}(y_{-},y_{+})
{\eta}_i(y_{-},y_{+}).
\end{eqnarray*}
For the free center of mass motion,
\[
{\eta}_i^{\star}(y_{-},y_{+}) {\eta}_i(y_{-},y_{+}) =
{\eta}_i^{\star}(y_{-}) {\eta}_i(y_{-}),
\]
so the integrals $\int_{-\infty}^{\infty} dy_{+} 
D\left(x,\frac{1}{2} (y_{+} \pm y_{-})\right)$ diverge.

We can not assume also that the dependence of the components on
the coordinates $x_{-}$ and $x_{+}$ is factorized, because for
a general form of $f(x_{-},x_{+})$ such factorization is simply
not valid. Even in the case of the purely Coulomb interaction
the factorization takes place only when the motion of the center
of mass is free. To prove that let us use for a moment the
following ansatz for the components ${\eta}_{+}$ and ${\chi}_{+}$:
\begin{eqnarray*}
{\eta}_{+}(x_{-},x_{+}) & = & {\eta}_{+}(x_{+}) {\eta}_{+}(x_{-}),\\
{\chi}_{+}(x_{-},x_{+}) & = & {\chi}_{+}(x_{+}) {\chi}_{+}(x_{-}).
\end{eqnarray*}
With $f={\phi}_{-}+E$, the equation (3.5a) gives
\[
{\eta}_{+}(x_{+}) = {\chi}_{+}(x_{+}),
\]
while the equation (3.5b) becomes
\[
\frac{1}{{\chi}_{+}(x_{+})} \frac{d^2{\chi}_{+}}{dx_{+}^2} =
\frac{f^2 {\chi}_{+}(x_{-}) -2fmc^2 {\eta}_{+}(x_{-})}
{-4{\hbar}^2c^2 {\chi}_{+}(x_{-})} .
\]
Since the left-hand side of this equation depends only on
$x_{+}$ and the right-hand one only on $x_{-}$ , both sides
must be equal to an arbitrary constant. Choosing the constant
as $(- P_{cm}^2/{{\hbar}^2})$, we get
\[
{\chi}_{+}(x_{+}) = \exp\left( \frac{i}{\hbar} P_{cm} 
x_{+} \right) ,
\]
i.e. the factor corresponding to the free motion of the
center of mass (if the constant is taken positive, say
$1/R_{cm}^2$, where $R_{cm}$ is a parameter of the dimension
of length, then we come to the factors $\exp\left( 
\pm x_{+}/R_{cm} \right)$
which diverge at positive or negative infinity and are
therefore unacceptable~).

The self-potentials are usually calculated by iteration
procedure. To lowest order of iteration we solve the spectrum
problem without the self-field terms. Then we substitute the
solution obtained into the expressions for the self-potentials,
calculate these potentials explicitly and use them in the next
order of iteration. Thus, to get finite expressions for the
self-potentials and to continue the iteration procedure we need
at the lowest order, i.e. in the Coulomb interaction case, a
general solution of the problem without the assumption of the
free motion of the center of mass.

\begin{center}

\section{DISCUSSION}

\end{center}

1. We have shown that the spectrum problem
for the two-body Hamiltonian in $(1+1)$-dimensional QED
reduces to the problem of solving a system of two
second-order partial differential equations~.
If the center of mass motion of the two-body system is
assumed to be free, then these equations govern only the
relative motion and take the form of
one-dimensional stationary Schr\"{o}dinger
type equations with energy-dependent potentials which
include the $\delta$-functional and inverted ocsillator
parts .

We have solved the problem in
the case of equal masses and the self-potentials
neglected, and derived the conditions determining the 
metastable energy levels. We have estimated the energies 
and widths of the metastable levels for large values of
mass. For the vanishing mass, neither stable nor
metastable levels exist, and the energy spectrum is
continuous.

2. Our consideration on the basis of the two-body equation
without the self-field terms does not change the result
of the single-particle Dirac equation approach, namely,
the nonexistence of hydrogenlike systems in $(1+1)$-dimensions.
However, the two-body equation, even with the self-interaction
neglected, provides essentially new details: for limited times
the particles can be confined in a metastable system
characterized by quasi-discrete energy levels. For large
values of the particle masses, the metastable system does
not decay for a long time, and its spectrum is close to
a discrete one.

To treat the problem completely it is necessary to take
into account in the two-body equation the self-potentials.
It is also of a principal importance to consider the
center of mass motion as a finite one, since for the free
motion case the self-potentials take infinitely large
values. This work is in progress.

\newpage

\begin{center}

{\bf APPENDIX}

\end{center}

\vspace{1 cm}

The function $G({\alpha},{\gamma};z)$ is defined as
\[
G({\alpha},{\gamma};z) = \frac{{\Gamma}(1-{\gamma})}{2{\pi}i}
\int_{{\rm C}_1} (1+ \frac{t}{z})^{-{\alpha}} t^{{\gamma}-1}
e^t dt,
\]
where the contour ${\rm C}_1$ comes from infinity
$({\rm Re}t \to -\infty)$, goes round the point $t=0$
and then returns to infinity (see Fig. 4.).

The asymptotic expansion of $G$ for $|z| \to \infty$ is
\[
G({\alpha},{\gamma};z) \approx 1 + \frac{{\alpha}{\gamma}}{z}
+ \frac{{\alpha}({\alpha}+1){\gamma}({\gamma}+1)}{2!z^2}
+ . . .     
\]
The relation between $G$ and its derivative with respect to
$z$ is
\[
G^{\prime}({\alpha},{\gamma};z) = - \frac{{\alpha}{\gamma}}{z^2}
G({\alpha}+1,{\gamma}+1;z).
\]
Other relations are
\begin{eqnarray*}
G({\alpha},{\gamma};z) & = & G({\alpha}+1,{\gamma};z) -
\frac{\gamma}{z} G({\alpha}+1,{\gamma}+1;z), \\
G({\alpha},{\gamma}+1;z) & = & \frac{\alpha}{z} G({\alpha}+1,
{\gamma}+1;z) + G({\alpha},{\gamma};z),
\end{eqnarray*}
and
\[
G^{\prime}({\alpha},{\gamma};z) =  \frac{\alpha}{z}
\left[ G({\alpha},{\gamma};z) - G({\alpha}+1,{\gamma};z) \right], 
\]
\[
G^{\prime}({\alpha},{\gamma};z) =  \frac{\gamma}{z}
\left[ G({\alpha}, {\gamma}+1;z) - G({\alpha},{\gamma};z) \right].
\]

\vspace{3 cm}

{\bf Acknowledgement}

\vspace{5 mm}

This research was financially supported by the Royal Society.

\newpage

\begin{center}

{\bf Figure Captions}

\end{center}

\begin{flushleft}

\vspace{5 mm}

Figure $1$\\

\vspace{2 mm}

The spectrum of the two-body system for $P_{cm}=0$;
(a) free motion; (b) with the Coulomb interaction.
The width of the metastable states is not shown.

\vspace{5 mm}

Figure $2$\\

\vspace{2 mm}

The form of the potential $V(x_{-})$ in the case
of equal masses and without the self-field terms
for $s>0$, i.e. for $e_1e_2>0$, $E>0$ or 
$e_1e_2<0$, $E<0$. Only the case $E^2 < \frac{\sqrt{3}}{2}
{\hbar}c|e_1e_2|$ is shown.

\vspace{5 mm}

Figure $3$\\

\vspace{2 mm}

The form of the potential $V(x_{-})$ in the case
of equal masses and without the self-field terms
for $s<0$, i.e. for $e_1e_2>0$, $E<0$ or
$e_1e_2<0$, $E>0$. Only the case $E^2 > \frac{\sqrt{3}}{2}
{\hbar}c|e_1e_2|$ is shown.

\vspace{5 mm}

Figure $4$\\

\vspace{2 mm}

The contour ${\rm C}_1$.

\end{flushleft}

\newpage

\setlength{\unitlength}{1cm}
\begin{center}
\begin{picture}(12,10)(-6,0)

{\thicklines
\put (-5,3){\line(1,0){3}}
\put (-5,7){\line(1,0){3}}
\put (0,3){\line(1,0){3}}
\put (0,7){\line(1,0){3}}
\put (0,3.4){\line(1,0){3}}
\put (0,4.9){\line(1,0){3}}
\put (0,5.1){\line(1,0){3}}
\put (0,6.6){\line(1,0){3}}

\multiput (-1.7,3)(.6,0){3}{\line(1,0){.3}}
\multiput (-1.7,7)(.6,0){3}{\line(1,0){.3}}
\multiput (3.3,3)(.6,0){2}{\line(1,0){.3}}
\multiput (3.3,7)(.6,0){2}{\line(1,0){.3}}}

\put (-3.4,.2){(a)}
\put (1.4,.2){(b)}
\put (4.5,7.2){$+2mc^2$}
\put (4.5,3.2){$-2mc^2$}

\put (0,7){\line(3,2){3}}
\put (0,7.2){\line(3,2){2.7}}
\put (0,7.4){\line(3,2){2.4}}
\put (0,7.6){\line(3,2){2.1}}
\put (0,7.8){\line(3,2){1.8}}
\put (0,8){\line(3,2){1.5}}
\put (0,8.2){\line(3,2){1.2}}
\put (0,8.4){\line(3,2){.9}}
\put (0,8.6){\line(3,2){.6}}
\put (0,8.8){\line(3,2){.3}}
\put (.3,7){\line(3,2){2.7}}
\put (.6,7){\line(3,2){2.4}}
\put (.9,7){\line(3,2){2.1}}
\put (1.2,7){\line(3,2){1.8}}
\put (1.5,7){\line(3,2){1.5}}
\put (1.8,7){\line(3,2){1.2}}
\put (2.1,7){\line(3,2){.9}}
\put (2.4,7){\line(3,2){.6}}
\put (2.7,7){\line(3,2){.3}}

\put (0,1){\line(3,2){3}}
\put (0,1.2){\line(3,2){2.7}}
\put (0,1.4){\line(3,2){2.4}}
\put (0,1.6){\line(3,2){2.1}}
\put (0,1.8){\line(3,2){1.8}}
\put (0,2){\line(3,2){1.5}}
\put (0,2.2){\line(3,2){1.2}}
\put (0,2.4){\line(3,2){.9}}
\put (0,2.6){\line(3,2){.6}}
\put (0,2.8){\line(3,2){.3}}
\put (.3,1){\line(3,2){2.7}}
\put (.6,1){\line(3,2){2.4}}
\put (.9,1){\line(3,2){2.1}}
\put (1.2,1){\line(3,2){1.8}}
\put (1.5,1){\line(3,2){1.5}}
\put (1.8,1){\line(3,2){1.2}}
\put (2.1,1){\line(3,2){.9}}
\put (2.4,1){\line(3,2){.6}}
\put (2.7,1){\line(3,2){.3}}

\put (-5,1){\line(3,2){3}}
\put (-5,1.2){\line(3,2){2.7}}
\put (-5,1.4){\line(3,2){2.4}}
\put (-5,1.6){\line(3,2){2.1}}
\put (-5,1.8){\line(3,2){1.8}}
\put (-5,2){\line(3,2){1.5}}
\put (-5,2.2){\line(3,2){1.2}}
\put (-5,2.4){\line(3,2){.9}}
\put (-5,2.6){\line(3,2){.6}}
\put (-5,2.8){\line(3,2){.3}}
\put (-4.7,1){\line(3,2){2.7}}
\put (-4.4,1){\line(3,2){2.4}}
\put (-4.1,1){\line(3,2){2.1}}
\put (-3.8,1){\line(3,2){1.8}}
\put (-3.5,1){\line(3,2){1.5}}
\put (-3.2,1){\line(3,2){1.2}}
\put (-2.9,1){\line(3,2){.9}}
\put (-2.6,1){\line(3,2){.6}}
\put (-2.3,1){\line(3,2){.3}}

\put (-5,7){\line(3,2){3}}
\put (-5,7.2){\line(3,2){2.7}}
\put (-5,7.4){\line(3,2){2.4}}
\put (-5,7.6){\line(3,2){2.1}}
\put (-5,7.8){\line(3,2){1.8}}
\put (-5,8){\line(3,2){1.5}}
\put (-5,8.2){\line(3,2){1.2}}
\put (-5,8.4){\line(3,2){.9}}
\put (-5,8.6){\line(3,2){.6}}
\put (-5,8.8){\line(3,2){.3}}
\put (-4.7,7){\line(3,2){2.7}}
\put (-4.4,7){\line(3,2){2.4}}
\put (-4.1,7){\line(3,2){2.1}}
\put (-3.8,7){\line(3,2){1.8}}
\put (-3.5,7){\line(3,2){1.5}}
\put (-3.2,7){\line(3,2){1.2}}
\put (-2.9,7){\line(3,2){.9}}
\put (-2.6,7){\line(3,2){.6}}
\put (-2.3,7){\line(3,2){.3}}

\end{picture}

\vspace{2 cm}

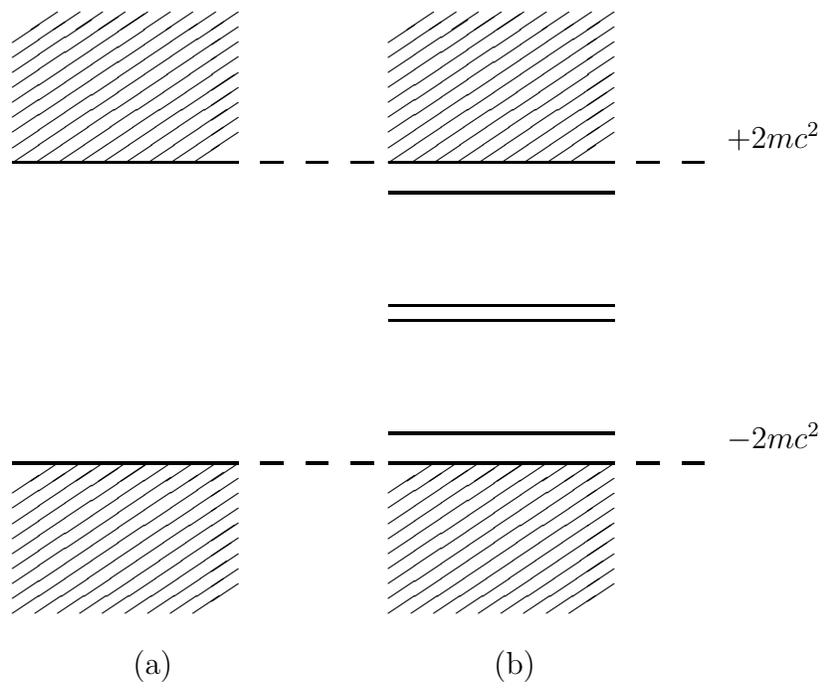
{\bf FIG. 1.}

\end{center}

\newpage

\setlength{\unitlength}{1cm}
\begin{center}
\begin{picture}(12,10)(-6,0)

\put (0,5){\line(0,1){3}}
\put (0,8){\line(0,1){1}}
\put (0,9){\vector(0,1){1}}
\put (0,5){\line(0,-1){3}}
\put (0,2){\line(0,-1){2}}
\put (-6,5){\line(1,0){3}}
\put (-3,5){\line(1,0){3}}
\put (3,5){\line(1,0){2}}
\put (5,5){\vector(1,0){1}}
\put (0,5){\line(1,0){3}}
\put (6,5.2){$x_{-}$}
\put (.2,10){$V$}

\put (-3,4.9){\makebox(0,0)[tr]{-s}}
\put (3,4.9){\makebox(0,0)[tl]{s}}
\put (-3,5){\circle*{.1}}
\put (3,5){\circle*{.1}}
\put (.2,7){$\tilde{V}_0$}
\put (0,6.8){\circle*{.1}}

{\thicklines
\put (-.1,6.8){\line(-1,-1){.4}}
\put (.1,6.8){\line(1,-1){.4}}
\put (-.6,6.3){\line(-3,-4){.3}}
\put (.6,6.3){\line(3,-4){.3}}
\put (-1,5.7){\line(-3,-5){.3}}
\put (1,5.7){\line(3,-5){.3}}
\put (-1.4,5){\line(-2,-5){.2}}
\put (1.4,5){\line(2,-5){.2}}
\put (-1.7,4.2){\line(-1,-3){.2}}
\put (1.7,4.2){\line(1,-3){.2}}
\put (-1.9,3.4){\line(-1,-6){.1}}
\put (1.9,3.4){\line(1,-6){.1}}

\multiput (-2,1)(0,.6){3}{\line(0,1){.4}}
\multiput (2,1)(0,.6){3}{\line(0,1){.4}}
\multiput (-.1,1.4)(0,.6){9}{\line(0,1){.4}}
\multiput (.1,1.4)(0,.6){9}{\line(0,1){.4}}

\put (0,1.2){\oval(.2,.4)[b]}}

\end{picture}

\vspace{2 cm}

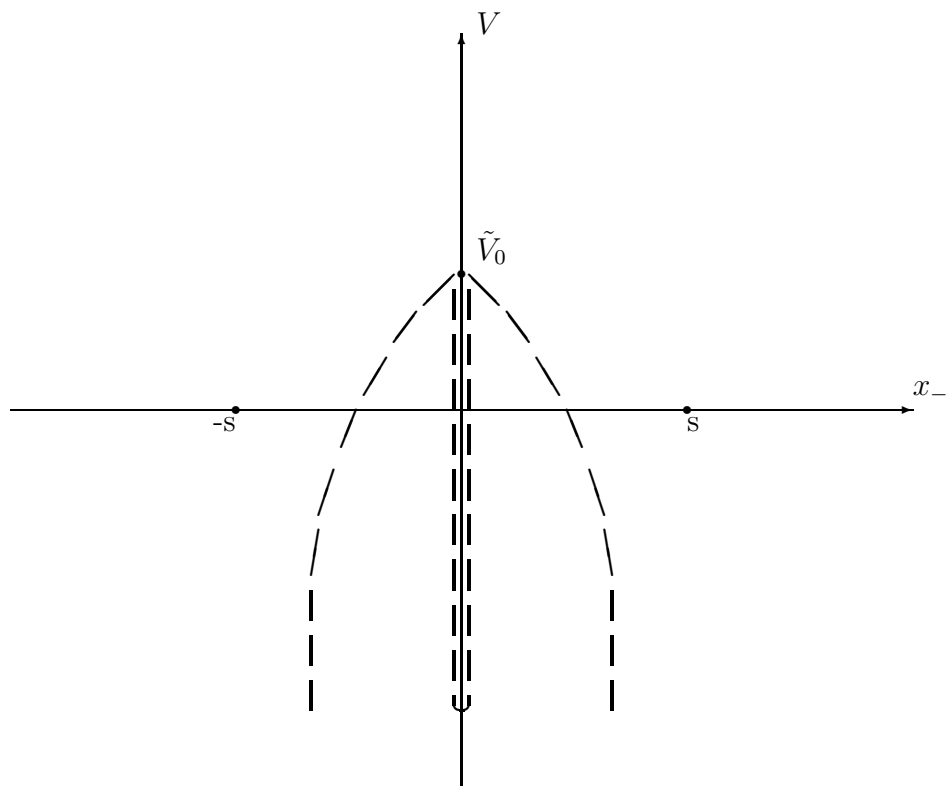
{\bf FIG. 2.}

\end{center}

\newpage

\setlength{\unitlength}{1cm}

\begin{center}
\begin{picture}(12,10)(-6,0)
\put (0,5){\line(0,1){3}}
\put (0,8){\line(0,1){1}}
\put (0,9){\vector(0,1){1}}
\put (0,5){\line(0,-1){3}}
\put (0,2){\line(0,-1){2}}
\put (-6,5){\line(1,0){3}}
\put (-3,5){\line(1,0){3}}
\put (0,5){\line(1,0){3}}
\put (3,5){\line(1,0){2}}
\put (5,5){\vector(1,0){1}}
\put (6,5.2){$x_{-}$}
\put (.2,10){$V$}

\put (-3,4.9){\makebox(0,0)[tr]{s}}
\put (3,4.9){\makebox(0,0)[tl]{-s}}
\put (-3,5){\circle*{.1}}
\put (3,5){\circle*{.1}}
\put (0,3.2){\circle*{.1}}
\put (.2,2.8){$\tilde{V}_0$}

{\thicklines
\put (0,8.8){\oval(.2,.4)[t]}
\multiput (.1,8.6)(0,-.6){9}{\line(0,-1){.4}}
\multiput (-.1,8.6)(0,-.6){9}{\line(0,-1){.4}}

\multiput (5.7,0)(0,.6){3}{\line(0,1){.4}}
\multiput (-5.7,0)(0,.6){3}{\line(0,1){.4}}
\multiput (2.9,9)(0,-.6){3}{\line(0,-1){.4}}
\multiput (3.1,9)(0,-.6){3}{\line(0,-1){.4}}
\multiput (-2.9,9)(0,-.6){3}{\line(0,-1){.4}}
\multiput (-3.1,9)(0,-.6){3}{\line(0,-1){.4}}
\put (2.9,7.2){\line(0,-1){.3}}
\put (-2.9,7.2){\line(0,-1){.3}}

\put (.1,3.2){\line(2,1){.4}}
\put (-.1,3.2){\line(-2,1){.4}}
\put (.8,3.6){\line(5,4){.4}}
\put (-.8,3.6){\line(-5,4){.4}}
\put (1.4,4.1){\line(1,1){.4}}
\put (-1.4,4.1){\line(-1,1){.4}}
\put (2,4.7){\line(4,5){.3}}
\put (-2,4.7){\line(-4,5){.3}}
\put (2.4,5.4){\line(1,2){.2}}
\put (-2.4,5.4){\line(-1,2){.2}}
\put (2.7,6.1){\line(1,3){.2}}
\put (-2.7,6.1){\line(-1,3){.2}}

\put (5.7,1.8){\line(-1,4){.1}}
\put (-5.7,1.8){\line(1,4){.1}}
\put (5.5,2.5){\line(-2,5){.2}}
\put (-5.5,2.5){\line(2,5){.2}}
\put (5.2,3.2){\line(-1,2){.2}}
\put (-5.2,3.2){\line(1,2){.2}}
\put (4.8,3.9){\line(-2,3){.3}}
\put (-4.8,3.9){\line(2,3){.3}}
\put (4.3,4.5){\line(-3,4){.3}}
\put (-4.3,4.5){\line(3,4){.3}}
\put (3.9,5.2){\line(-3,5){.3}}
\put (-3.9,5.2){\line(3,5){.3}}
\put (3.5,5.9){\line(-2,5){.2}}
\put (-3.5,5.9){\line(2,5){.2}}
\put (3.2,6.6){\line(-1,6){.1}}
\put (-3.2,6.6){\line(1,6){.1}}}

\end{picture}

\vspace{2 cm}

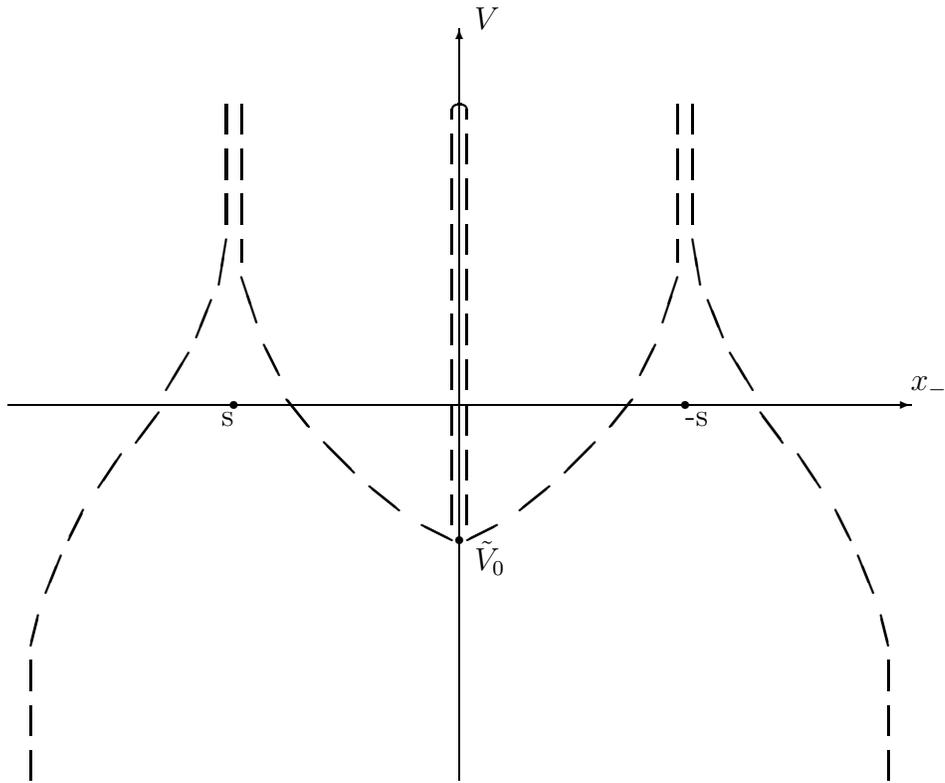
{\bf FIG. 3.}

\end{center}

\newpage

\setlength{\unitlength}{1 cm}
\begin{center}
\begin{picture}(12,10)(-6,0)
{\thicklines
\put (-4,4.5){\line(1,0){3}}
\put (-1,4.5){\vector(1,0){1}}
\put (0,4.5){\line(1,0){3}}
\put (-4,5.5){\line(1,0){4}}
\put (3,5.5){\line(-1,0){2}}
\put (1,5.5){\vector(-1,0){1}}
\put (3,5){\oval(2,1)[r]}}

\put (3,5){\circle*{.1}} 
\put (4.1,4){${\rm C}_1$}
\put (1.8,4.9){$t=0$}
\end{picture}

\vspace{2 cm}

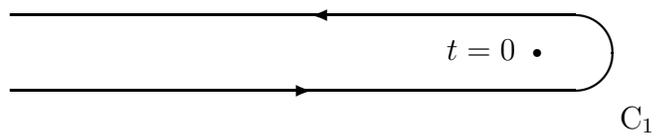
{\bf FIG. 4.}

\end{center}

\end{document}